\documentclass[epj]{svjour}
\usepackage[english]{babel} 
\usepackage{graphicx}
\usepackage{amsmath,amsfonts,amssymb} 
 \usepackage{color} 

\begin{document}
\title{ Desynchrony and synchronisation underpinning sleep-wake cycles}

\author{
Dmitry E. Postnov\inst{1}\thanks{postnov@info.sgu.ru}  \and  
Ksenia O. Merkulova\inst{1} \and
Svetlana Postnova \inst{2,3,4}
}
 
\institute{ Saratov State University, 83 Astrakhanskaya street, Saratov 410012, Russia \and School of Physics, University of Sydney, NSW 2006, Australia \and The University of Sydney Nano Institute (Sydney Nano), The University of Sydney, NSW 2006, Australia \and Charles Perkins Centre, The University of Sydney, NSW 2006, Australia}
\date{Received: date / Revised version: date}
 \abstract{ 
{~\\
\bf Objectives:} This paper studies mechanisms of synchronisation and loss of synchrony among the three key oscillatory processes controlling sleep-wake cycles in the human brain: the 24 h circadian oscillator, the homeostatic sleep drive, and the environmental light-dark cycle. Synchronisation of these three rhythms promotes sleep and brain clearance and is critical for human health. Their desynchrony, on the other hand, is associated with impaired performance and disease development, including cancer, cardiovascular disease and mental disorders.\\
{\bf Methods:} A biophysical model of arousal dynamics simulating sleep-wake cycles and circadian rhythms is used as the study system. It is based on established neurobiological mechanisms controlling sleep-wake transitions and incorporates the three oscillatory processes. Nonlinear dynamics methods and synchronisation theory are used to numerically investigate model dynamics under conditions that are not easily achievable in experiments. The role of homeostatic brain clearance rate in synchronisation is investigated and selective turning on and off of coupling strengths between the oscillators allows us to determine their role in oscillators' dynamics.\\
{\bf Results:} We find that the default state of the model corresponds to the endogenous homeostatic period that is far from $\sim$24 h rhythm of the circadian and light-dark cycles. Combined action of light and circadian oscillator on the homeostatic rhythm is required to achieve the typical sleep-wake pattern that is observed in young healthy people. Change of homeostatic clearance rate is found to induce two types of desynchronisation: (i) fast clearance rates $\tau_H<58.1$ h desynchronise the homeostatic oscillator from the circadian, while the circadian rhythm remains entrained to the light-dark cycle, and (ii) slow clearance rates $\tau_H>69$ h maintain synchronisation between the homeostratic and circadian oscillators, but the period of both is different from that of the light-dark cycle. Between these regimes, all three rhythms are synchronised under the studied conditions. The model predicts that the system is highly sensitive to external inputs to the neuronal populations of the sleep-wake switch, which affect the endogenous period of the homeostatic oscillator and can lead to complete loss of sleep. \\ 
{\bf Conclusions:} Model dynamics show that loss of synchronisation, which is traditionally ascribed to impairment of the circadian oscillator, can be caused by changes in the homeostatic clearance rate of the brain or external input to the neuronal populations of the sleep-wake switch. This has significant implications for understanding individual variability in sleep-wake patterns and in mechanisms of sleep and circadian disorders, indicating that both the homeostatic and circadian mechanisms can be responsible for the same clinical or behavioural presentation of a disease.
}

\maketitle
 
\section{Introduction}
 
Sleep is a crucial time for clearance of toxic neuro-metabolites from the brain \cite{xie2013sleep}. This process is driven by the recently discovered glymphatic system - a brain-wide perivascular passageway that transports waste products out of the brain and in to the cerebrospinal fluid \cite{Iliff2012}. Glymphatic system is more active during sleep than wake \cite{xie2013sleep} and its activation has been linked to the slow wave activity (SWA) observed in the electroencephalography (EEG) during non-rapid eye movement (NREM) sleep \cite{fultz2019coupled}. In line with these findings, sleep deprivation and selective suppression of slow wave activity during sleep led to accumulation of waste products in the brain \cite{Ju2017,Kang2009}. Like other physiological functions, the process of brain clearance is under circadian control. This is realised both indirectly via circadian regulation of sleep time and quality \cite{Golombek10,Dijk95,Dijk1997} and directly via circadian distribution of CSF in the brain supported by aquaporin-4 channels dynamics in astrocytes \cite{hablitz2020circadian}. However, the overall complex system responsible for the clearance and accumulation of neuro-toxic waste products during sleep-wake cycles is yet to be fully understood \cite{fultz2019coupled,ding2016changes,ingiosi2020role,hablitz2020circadian}.

The timing of sleep is under circadian control with highest NREM SWA power, and potentially fastest clearance, achieved when sleep appears during circadian rest phase (night for humans) \cite{Dijk1997}. The daily sleep-wake pattern depends on stable phase relationship between sleep homeostasis, circadian oscillator, and the environmental inputs (e.g., the light-dark cycle) \cite{Golombek10}. Sleep homeostasis reflects the sleep need, which increases during wakefulness and declines during sleep \cite{borbely1982two}. The power of SWA in NREM sleep EEG is the current 'gold standard' marker of sleep homeostasis \cite{Borbely81}. The exact mechanisms of sleep homeostasis are unknown, but thought to be associated with accumulation and clearance of the somnogens, or toxic waste products, and changes in synaptic connectivity and astroglial calcium signalling in the brain \cite{porkka00,Datta10,Allada17,ingiosi2020role}. 

The central circadian oscillator in the suprachiasmatic nucleus (SCN) of the hypothalamus controls the 24 h periodicity of the sleep-wake cycles. It promotes wakefulness by counteracting the homeostatic need for sleep during daytime and enables consolidated sleep episodes during night in humans \cite{borbely1982two,Dijk95,Golombek10}. The phase of the circadian oscillator, in turn, is affected by timing and intensity of environmental and behavioural factors, such as light-dark cycle, meals, and physical activity \cite{Golombek10,Kalsbeek14,Youngstedt19}. Phase alignment and synchronisation of these three rhythms: sleep homeostasis, circadian oscillator, and external drives, are critical for optimal sleep and health.

Circadian misalignment is observed when sleep-wake cycles, circadian oscillator and/or environmental factors are out of sync with each other. Such misalignment of rhythms became common in the modern society with jetlag and shiftwork being an integral part of life for many people \cite{Finger2020}. In these cases, misalignment is caused by changes in environmental factors and behaviour, which lead to desycnhronisation between the circadian oscillator and the sleep homeostat \cite{Golombek10,Finger2020}. A rare example of circadian misalignment is spontaneous internal desynchrony, where long-term absence of environmental inputs leads to desynchronisation of sleep and circadian oscillator - such conditions do not appear in the real life except for some blind individuals. In this case, the circadian oscillator functions at $\sim$24 h period, while sleep appears at much shorter (12-20 h) or longer (28-68 h) intervals \cite{Wever79,Phillips11,Aschoff67,Aschoff71,Gleit13}. Circadian misalignment is a known risk factor for disease development, including metabolic, cardiovascular and neurological diseases \cite{Kalsbeek14,Baron14}. During circadian misalignment sleep is disturbed and appears at sub-optimal circadian phases. This likely leads to disturbances in brain drainage and clearance, which can result in long-term accumulation of toxins and development of disease \cite{Ju2017,Kang2009,Holth19}. 

 Mathematical models were developed to capture the interactions between sleep and circadian rhythms (reviewed in \cite{Abel20,Postnova19}). These were successful in simulating normal sleep, effects of sleep deprivation on sleep homeostasis and recovery, sleep patterns of different mammals, and alertness dynamics, among other phenomena \cite{Abel20,Postnova19}. The model of arousal dynamics \cite{Phillips11,postnova2016sleep,Abeysuriya18,Tekieh2020}, in particular, focused on the interaction between the sleep-wake cycles, circadian oscillator and the external driving force. It was tested against both laboratory and real-world experimental data and, in addition to the above phenomena, reproduced circadian misalignment dynamics observed during shiftwork, jetlag, spontaneous internal desycnhrony and forced desynchrony protocols \cite{Phillips11,Abeysuriya18,postnova2016sleep,Gordon2018}. However, physical mechanisms of de-synchronisation in this model were not yet fully characterised. 
 
The model of arousal dynamics is composed of two coupled oscillators of different types (sleep homeostatic one  and circadian oscillator) with a external, usually periodic, driving forces. Such systems have been extensively studied in nonlinear dynamics, synchronisation theory \cite{pikovsky2003synchronization,balanov2008synchronization}, which explains mechanisms underlying synchronisation in different types of oscillating systems and provides mathematical tools for investigation of new systems.

In this study we use nonlinear dynamics tools to (i) investigate synchronisation mechanisms in the model of arousal dynamics, (ii) compare these mechanisms to classical models of synchronisation, and (iii) study the role of homeostatic clearance of the brain during sleep on synchronisation and sleep-wake patterns. The paper is structured as follows. In the Methods section, we describe model equations and simplify the model to convert it to a more conventional form which was extensively studied in synchronisation theory.  This will allow us to investigate sensitivity of the model to different parameters and compare synchronisation mechanisms in the model with those already established for other oscillators. In the Results section, we investigate model's synchronisation properties under conditions of different coupling configurations, form of external driving force (light-dark cycle), and varied properties of the homeostatic oscillator. Finally, in the last section, we discuss implications of our findings for physiology and modelling of sleep-wake cycles and for pathophysiology of circadian and sleep disorders.  
 
\section{Methods}
We focus on the basic version of the model of arousal dynamics \cite{postnova2016sleep} as it contains all components relevant to study synchronisation but excludes improvements and updates that do not affect on synchronisation. Later model extensions introducing dynamics of alertness \cite{Postnova18}, melatonin \cite{Abeysuriya18}, and effects of light spectrum \cite{Tekieh2020} do not change the core oscillating processes in the model and will follow the same synchronisation mechanisms as the original model. To enable comparison to well-studied oscillator models we reduce the number of equations in the homeostatic oscillator and simplify coupling between the oscillators. By doing so we demonstrate that the homeostatic oscillator in the model can be represented as the Fitzhugh-Nagumo (FHN) model, which is a well-known model in synchronisation theory and nonlinear physics in general \cite{fitzhugh1961impulses}.

\subsection{Model of arousal dynamics}
\begin {figure} [htbp]
\begin {center}
 	\includegraphics [width=0.8\textwidth]{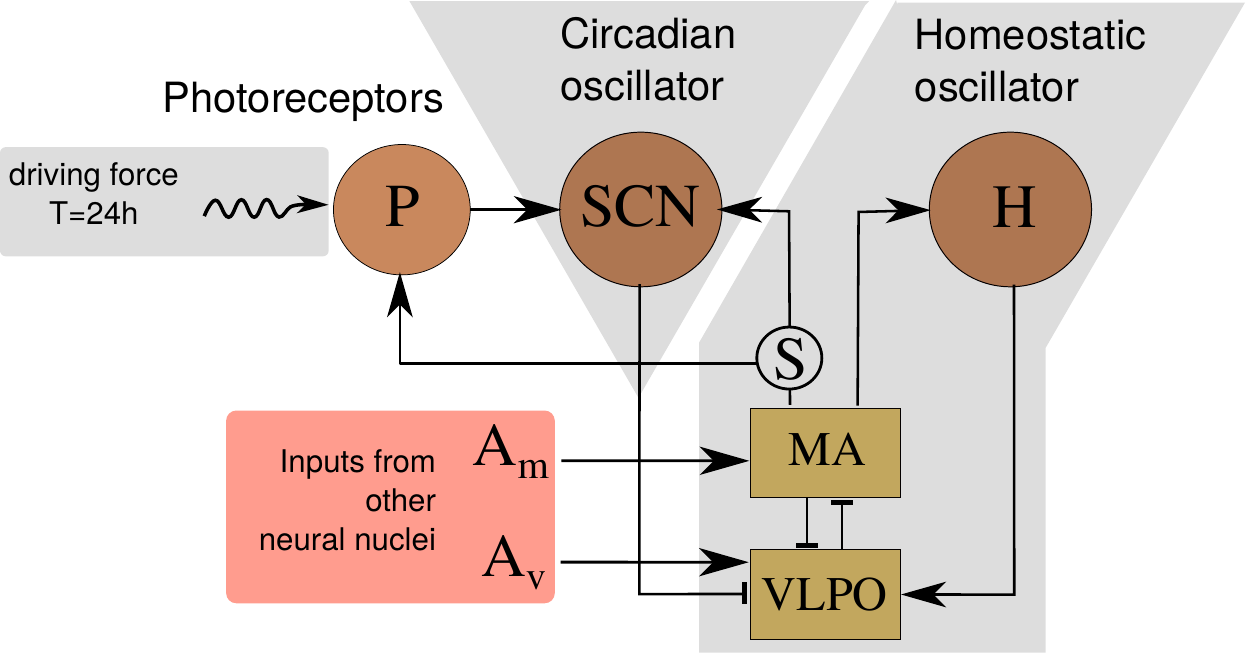}
\caption{ Schematic of the model of arousal dynamics adapted from \cite{postnova2016sleep}. The two oscillators and the driving force (the light-dark cycle) are highlighted with the grey background. Inputs from other neuronal populations, $A_m$ and $A_v$ are highlighted with the pink background. The notations are: P - photoreceptors, SCN - suprachiasmatic nucleus of the hypothalamus, $H$ - homeostatic drive, MA - monoaminergic nuclei of the hypothalamus and brainstem, VLPO - ventrolateral preoptic nucleus of the hypothalamus, $S$ - arousal state, which can be either sleep or wake. Arrows indicate excitatory connections and bar-headed lines - inhibitory.}
\label{schematic}
\end {center}
\end {figure}

The model of arousal dynamics \cite{postnova2016sleep} is based on a combination of the model of ascending arousal system in the brain regulating the sleep-wake switch \cite{phillips2007quantitative} and the dynamic circadian oscillator driven by light \cite{StHilaire07,Kronauer07}. Schematic of the key model components is shown in Fig.~\ref{schematic}. The light-dark cycle, i.e. the driving force, acts on the photoreceptors, P, in the eye. These, in turn, send an input to the circadian oscillator in the suprachiasmatic nucleus, SCN, of the hypothalamus, which has its own endogenous period of $\sim$24 h. The homeostatic oscillator is composed of the homeostatic drive, $H$, and the two mutually inhibitory neuronal populations: the sleep-active ventrolateral preoptic nucleus of the hypothalamus, VLPO, and the wake-active monoaminergic nuclei, MA, of the hypothalamus and brainstem \cite{saper2010sleep}. VLPO receives a sleep-promoting excitatory input from H and a wake-promoting inhibitory input from the SCN. MA controls the accumulation and decay of $H$ and determines the states of wake and sleep, $S$, which gate the light input to the brain and modulate the dynamics of the SCN through non-photic coupling. Parameters, $A_m$ and $A_v$ represent inputs from other neuronal populations to the MA and VLPO, respectively, including those from the orexinergic and cholinergic nuclei. In the original model, these are assumed to be constant.

\subsubsection{Homeostatic oscillator}
The homeostatic oscillator is composed of the MA, VLPO and $H$ and is described by the following equations:
\begin{align}
&\tau_v \frac{dV_v}{dt} =  \nu_{vm} Q(V_m) - V_v +  \nu_{vH}H  + A_v  +\nu_{vC}C(X,Y), \label{eqVv} \\ 
&\tau_m \frac{dV_m}{dt} =  \nu_{mv} Q(V_v)  - V_m +  A_m,  \label{eqVm}  \\ 
&\tau_H \frac{dH}{dt}   =  \nu_{Hm} Q(V_m)  - H,  \label{eqH}
\end{align}
where $V_v$ and $V_m$ are the mean voltages of the VLPO and MA populations, $\tau$ are the time constants of respective variables, $\nu_{ij}$ are the coupling strengths from model component $j$ to $i$. The states of sleep and wake, $S$, are determined from the dynamics of $V_m$: wake is registered when $V_m$ is above the threshold value $V_{th}$ ($V_m>V_{th}$) and sleep when $V_m\leq V_{th}$. 

The mean firing rate $Q$ is given by a sigmoid function of the mean voltage of a respective neuronal population
\begin{align}
&Q(V)=\frac{Q_{max}}{1 + e^{(\Theta-V)/\sigma '}}, \label{Q} 
\end{align}
where $Q_{max}$ is the maximum mean firing rate, $\Theta$ is half-activation threshold, and $\sigma '\pi/\sqrt{3}$ is the standard deviation of the threshold. Circadian input to the homeostatic oscillator is given by $\nu_{vC}C(X,Y)$, which is described in Sec.~\ref{sec_coupling_all}. 
The default parameter values are: $\tau_v = 50$ s, $\tau_m = 50$ s, $\tau_H = 59\times3600$ s, $\nu_{vm} =  -2.1$ mV, $\nu_{mv} =  -1.8$ mV, $\nu_{Hm} =   4.57$ s, $\nu_{vh} = 1.0$, $\nu_{vc} = -0.5$ mV, $A_v = -10.3$ mV, $A_m = 1.3$ mV; $Q_{max}=100$ Hz, $\Theta=10$ mV, $\sigma ' =3$ mV. 

\subsubsection{Circadian oscillator}
The circadian oscillator follows Van der Pol equations and is represented by two circadian variables X,Y 
\begin{align} 
&\tau_x \frac{dX}{dt}    =  Y   + \gamma \left( \frac{1}{3}X + \frac{4}{3}X^3 - \frac{256}{105}X^7 \right)
 +C_{Xn} + C_{Xp} ,  \label{eqX} \\ 
&\tau_y \frac{dY}{dt}    = -\left(\frac{\delta}{\tau_c}\right)^2X  + C_{Yp},\label{eqY}
\end{align}
where $\tau_x = \tau_y$ scale the oscillator period to 24 hours, $\gamma$ is the stiffness, $\tau_c$ is the endogenous circadian period, and $\delta$ scales the period for consistency with experimental data \cite{StHilaire07}. Input functions $C_{Xp}, C_{Yp}$ and $C_{Xn}$ describe photic (light-dependent) and non-photic (sleep-wake state-dependent) influences on the circadian oscillator.  

The default parameter values are: $\tau_c  = 24.2\times3600$ s, $\tau_x  = 24.0\times3600/(2\pi)$ s,  $\tau_y  = 24.0\times3600/(2\pi)$ s, $\gamma =   0.13$, $\delta = 24.2\times3600/0.99729$ s,

\subsubsection{Coupling terms} \label{sec_coupling_all}
\subsubsection*{Circadian coupling (circadian $\to$ homeostatic)}
The influence of the circadian oscillator on the homeostatic one is introduced with the term $\nu_{vC}C(X,Y)$ in Eq.(\ref{eqVv}) where $\nu_{vC}$ is the circadian coupling strength, $X$ and $Y$ are the variables of the circadian oscillator, and $c_1, c_2, c_3$ are the weighting parameters adjusting the shape of the circadian drive. The circadian drive $C(X,Y)$ is given as a nonlinear function of the circadian variables
\begin{align}
&C(X,Y) = 0.05X + \left( \frac{c_1X+c_2Y +c_3}{X+2}\right)^2. \label{eqC}
\end{align}
The default parameter values are $\nu_{vc} = -0.5$ mV, $c_1= 0.095$, $c_2=0.676 $, $c_3=1.136$

\subsubsection*{Non-photic coupling (homeostatic $\to$ circadian)}
The non-photic coupling, $C_{Xn}$ simulates modulation of the circadian oscillator depending on model's arousal state. Effectively this provides coupling from the homeostatic to the circadian oscillator:
\begin{align} 
&C_{Xn} = \nu_{Xn}  \left( \frac{1}{3} -S \right)(1-tanh(rX)), \label{eqCXn}\\
&S = U(V_m-V_{th}), \label{eqS}
\end{align}
where $\nu_{Xn}$ is the non-photic coupling strength, and $r$ regulates the timing of the non-photic effects. The state function, $S$, takes a value of unit during wake (when $V_m > V_{th}$) and zero during sleep ($V_m\leq V{th}$). Here, $U(x)$ is the unit function, $U(x)=1$ if $x>0$, and $U(x)=0$ otherwise, $V_{th}$ is the voltage sleep threshold.

The default parameter values are: $\nu_{Xn} = 0.032$, $r=10$, $V_{th}=-2$ mV. 

\subsubsection*{Photic coupling (light-dark cycle $\to$ circadian)}
Input from the external driving force, the light-dark cycle, to the circadian oscillator is given by the photic coupling functions $C_{Xp}$ and $C_{Yp}$ to each of the circadian variables, respectively. These depend on the fraction of activated photoreceptors $P$ in the eye and photopic illuminance input $I$:
\begin{align} 
&C_{Xp} = \nu_{Xp} \alpha_I(1-P)(1-\epsilon X)(1-\epsilon Y), \label{eqCXp}\\
&C_{Yp} = \alpha_I(1-P)(1-\epsilon X)(1-\epsilon Y) (\nu_{YY}Y - \nu_{YX}X), \label{CYp}\\
&\tau_p\frac{P}{dt} = \alpha_I(1-P) - \beta P, \label{eqP} \\
&\alpha_I =   \alpha_0 S\frac{I}{I+I_1}\sqrt{\frac{I}{I_0}}. \label{eqalpha}
\end{align}
Parameters $\nu_{ij}$ control coupling strength from $j$ to $i$, $\epsilon$ modulates sensitivity of the photic drive to the circadian variables, and $\tau_p$ is the time constant of receptor activation. The fraction of photoreceptors ready to be activated is $(1-P)$, which are converted to active state with the rate $\alpha_I$ and converted from active to ready at rate $\beta$ . Function $\alpha_I$ introduces effects of light $I$ on activation of photoreceptors and is gated to zero during sleep ($S=0$). Parameters $\alpha_0, I_0, I_1$ adjust the effects of light on activity of photoreceptors. 

We simulate 12/12 light-dark cycle using the unit function:
\begin{align} 
&I = I_{ext} U \left( \sin{ \left( \frac{2\pi}{24}(t-8.0) \right)} \right), \label{eqI}
\end{align}
where $I_{ext}$ is constant, and $t-8.0$ offsets the light phase to be from 08:00 to 20:00 and dark otherwise.  

The default parameter values are: $\alpha_0 =0.1/60.0$ s$^{-1}$,  $I_1 = 100$ lx, $I_0  = 9500$ lx, $\epsilon= 0.4$, $\nu_{Xp} = 37*60$ s, $\nu_{YY} = 12.33*60$ s, $\nu_{YX} = 20.35*60$ s,  $\tau_p  = 1$ s,   $\beta = 0.007/60$ s$^{-1}$, $I_{ext}$ is chosen depending on a simulated protocol. 
 
\subsection{Model reduction} 
\subsubsection{Simplified homeostatic oscillator matches Fitzhugh Nagumo model}
To enable synchronisation analysis of the model we first investigate if the homeostatic oscillator can be simplified to a lower-dimensional form. The time constants for the MA and VLPO are much smaller than for the homeostatic drive: $\tau_m$, $\tau_v$ $\ll$  $\tau_H$. This means that the homeostatic oscillator is a fast-slow oscillator with the period driven by the slow time constant and the slow movement on the limit cycle. Since the equation for $V_v$ includes coupling terms while the one for $V_m$ does not, we set $\tau_m \rightarrow 0$ and assume that the dynamics for $V_m$ are instantaneous, so $\tau_m dV_m/dt=0$ . Equation (\ref{eqVm}) can then be re-written as  
\begin{align}
& V_m=\nu_{mv} Q(V_v) +  A_m. 
\end{align}
Substituting $V_m$ to the equations for $V_v$ and $H$ we get 
 
\begin{align}
&\tau_v \frac{dV_v}{dt} =  \nu_{vm} Q(\nu_{mv} Q(V_v)+  A_m) - V_v +  \nu_{vH}H  + A_v  +\nu_{vC}C,  \label{rAAm_1}\\ 
&\tau_H \frac{dH}{dt}   =  \nu_{Hm} Q(\nu_{mv} Q(V_v)+  A_m)  - H.  \label{rAAm_2}
\end{align}

Oscillator described in Eqs (\ref{rAAm_1}),(\ref{rAAm_2}) is a 2D oscillator, which is a simplified version of the 3D oscillator in Eqs (\ref{eqVv})-(\ref{eqH}). 

To characterise the phase plane of the 2D homeostatic oscillator independent of the circadian influence we set $\nu_{vC}=0$. In this case,  we can find the $V_v$- and $H$-nullclines, which are
\begin{align}
 &V_v - nullcline      &H= \frac{1}{ \nu_{vH}}\{ V_v -  \nu_{vm}Q(\nu_{mv} Q(V_v)+  A_m)-A_v \}, \\
 &H - nullcline    &H= \nu_{Hm}Q(\nu_{mv} Q(V_v)+ A_m).
\end{align}

\begin{figure}[htb]
\begin{center}
\includegraphics[width=0.8\textwidth]{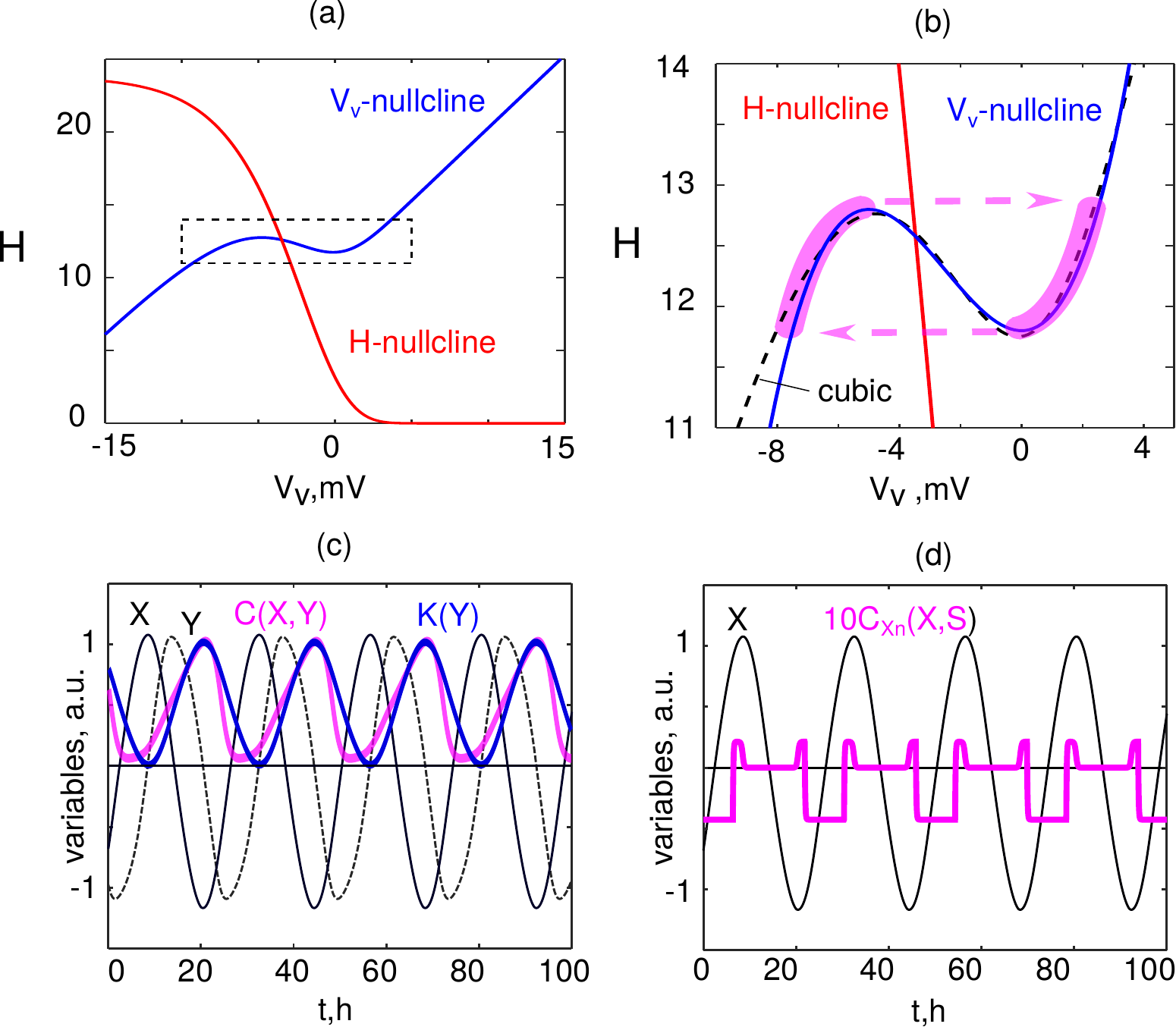}
\caption{Dynamical features of the homeostatic oscillator and coupling between the circadian and homeostatic oscillators. (a) The $H$- and $V_v$-nullclines of the homeostatic oscillator and (b) is the zoomed-in area critical for the self-sustained oscillations that is highlighted with a dashed rectangle in (a). Dashed arrows show the fast segments of the limit cycle and the direction of variables change. Slow segments are highlighted in magenta. The polynomial fitting for the $V_v$-nullcline  is denoted as 'cubic', $V_{cubic}= 0.75(z^3/3 -z) +12.3$, where $z=0.6V_v +1$. (c) Time-dependence of the variables relevant for the circadian input $\nu_{vC}C(X,Y)$ acting on the homeostatic oscillator. The circadian drive is shown in magenta, and its approximation with $K(Y)$ is shown in blue. 
(d) The non-photic coupling $C_{Xn}(X,S)$ acting on the circadian oscillator. Factor of 10 is used for better visualisation alongside $X$   variable.}
\label{nulls} 
\end{center}
\end{figure}

Figure ~\ref{nulls}(a) shows the nullclines on the $H$ vs. $V_v$ plane and highlights the range responsible for the self-sustained oscillations, i.e., the limit cycle in Fig. ~\ref{nulls}(b). In the range of self-sustained oscillations, the $H$-nullcline can be approximated with a line and the shape of the $V_v$-nullcline can be described with a cubic parabola as shown with a dashed line in Fig. ~\ref{nulls}(b) -  these are similar to nullclines of the FitzHugh-Nagumo (FHN) model \cite{fitzhugh1961impulses,izhikevich2007dynamical}. 

The mathematical form of the homeostatic oscillator is also identical to that of the FHN model. By making the following substitutions
\begin{align}
& v=V_v, \;\; h=-H, \;\; \gamma_h=1/\tau_H, \nonumber
 \end{align}
and approximating the $H$-nullcline in Fig.~\ref{nulls}(b) with a line having a slope $a$ and offset $b$
\begin{align}
 &\nu_{Hm} Q(\nu_{mv} Q(V_v)+  A_m)/\tau_H = a v + b,
\end{align}
we show that the 2D homeostatic oscillator in Eqs~(\ref{rAAm_1}),(\ref{rAAm_2}) becomes
 \begin{align}
\label{rAAm_11}
\varepsilon_v&\frac{dv}{dt} =  F(v)   -h; \;\;\;\;   
\frac{dh}{dt}   =  a v+ \gamma_h h + b,  
\end{align}
where $\varepsilon_v =\tau_v/\nu_{vH}$, and $F(v)=(\nu_{vm} Q(\nu_{mv} Q(v)+  A_m) - v)/(\nu_{vH})$. The system of equations in (\ref{rAAm_11}) exactly matches FHN model, which is considered one of the classical models in synchronisation theory and its oscillatory behaviour is well characterised \cite{balanov2008synchronization}. For example, the nullcine positions in Fig.~\ref{nulls}(b) correspond to self-sustained oscillations around a steady state, which is located at the crossing of the two nullclines. The oscillator's period is determined by the length of the slow segments on the $V_v$-nullcline (same as $v$-nullcline) (highlighted in magenta). The time spent by the model in the fast segments, shown with dashed arrows have only a minor contribution to the period of oscillations but exact position of the fast segments determines the lengths of the slow ones. Changes in model parameters can lead to disappearence of self-sustained oscillations, which happens when the crossing of the two nullclines moves to either one of the parabola extrema. This representation of the homeostatic oscillator is useful for understanding its synchronisation properties, which we use in the analysis throughout this study. %
 
\subsubsection{Coupling between the homeostatic and circadian oscillators} \label{sec_coupling}
Circadian oscillator acts on the homeostatic one via the term $\nu_{vC}C(X,Y)$ in Eqs~(\ref{eqVv}) and (\ref{rAAm_1}), where circadian drive $C(X,Y)$ is a nonlinear function of the circadian variables, Eq.~(\ref{eqC}).This means that the coupling is nonlinear and that it may lead to appearance of new oscillation frequencies dependent on the degree of the polynomial. In this work, however, we focus on 1:1 synchronisation between the homeostatic and circadian oscillators, so the harmonics introduced by the nonlinear coupling are less relevant. This allows us to approximate the coupling term, $C(X,Y)$ with a simpler linear function $K(Y)$ dependent only on one circadian variable.
\begin{align}
\label{approx}
& C(X,Y)\approx K(Y)        &K(Y) = c_4(1-Y),
\end{align}
where $c_4 \approx0.47$.
 
Figure \ref{nulls}(c) shows comparison of $C(X,Y)$ and $K(Y)$. Function $K(Y)$ demonstrates maximum wake-promoting circadian signal at the same time as $C(X,Y)$ (maxima of both functions). There is discrepancy in the timing of the minimum wake-promoting signal, but this should not significantly affect frequency synchronisation of the oscillators. Frequency/period synchronisation is expected to have similar properties when $K(Y)$ or $C(X,Y)$ but the shape of the signal will affect phase synchronisation. This is why $C(X,Y)$ is needed to reproduce nuances of the multitude of experimentally observed sleep phenomena as shown in the model of arousal dynamics.

The approximation allows us to separate the oscillatory and constant components in the coupling. The former is responsible for synchronisation of rhythms and the latter modulates the constant input to the VLPO and can thus be added to $A_v$ which affects the period of the homeostatic oscillator. Thus change of $\nu_{vC}$ has two effects: it affects the strength of the circadian action on the homeostatic oscillator and it modulates the homeostatic period, which both need to be taken into consideration when analysing model dynamics.   

The non-photic coupling from the homeostatic oscillator to the circadian is shown in Fig.~\ref{nulls}(d) and defined in Eqs~(\ref{eqCXn}),(\ref{eqS}). This coupling is weak compared to the magnitude of the circadian variables (note the factor of 10 for visualisation in Fig.~\ref{nulls}(d)) and is weaker than the coupling from the circadian oscillator to the homeostatic. Importantly, the non-photic term is negative during sleep (coinciding with increasing part of the $X$ variable), which means that the coupling is likely to slow down the circadian oscillator. 

The photic coupling term describes the effects of the driving light-dark force on the circadian oscillator and is defined by the functions $C_{Xp}$ and $C_{Yp}$ which, in turn, depend on several parameters and variables as described in Eqs (\ref{CYp}),(\ref{eqCXp}). Due to its complexity, the effects of this coupling on synchronisation need to be assessed numerically.

\subsection{Simulation protocols}

To collect information about the   homeostatic and circadian periods, the simulation was run  for 150 days of model time, and the last 100 days were used to calculate $ T_S $ and $ T_C $.
AS the starting points  $ T_C $ and $ T_S $  calculation, conditions $ Y (t) = 0 $ and $ V_v = V_ {th} $ were used, respectively. 
Maps of periods on the plane of two parameters were calculated by means of parallel computing, thus, each combination of parameters from the 400 $\times $ 100 matrix was checked.

\section{Results} 
\subsection{Endogenous homeostatic period is different from the $\sim$24 h circadian rhythm}
 
To investigate synchronisation properties of the homeostatic oscillator under influence of the circadian signal we calculate Arnold tongues diagram at varied strength of the circadian coupling, $\nu_{vC}$, but zero non-photic and photic coupling acting on the circadian oscillator ($I_{ext}=0$ lx, $\nu_{Xn}=0$). This approach is widely used in synchronisation theory \cite{pikovsky2003synchronization,balanov2008synchronization} and allows us to visualise synchronous states depending on the endogenous period of the homeostatic oscillator, controlled by $\tau_H$, and the strength of oscillatory coupling acting on it, $\nu_{vC}$. The endogenous period of the homeostatic oscillator is controlled by its slowest variable, $H$, whose rate of change is determined by the time constant $\tau_H$. Indeed, if $\nu_{vC}=0$, the period of the homeostatic oscillator can be approximated as $T_S = 1.4 + 0.25\tau_H$.  

\begin {figure} [htbp]
\begin {center}
 
\includegraphics [width=0.95\textwidth ]{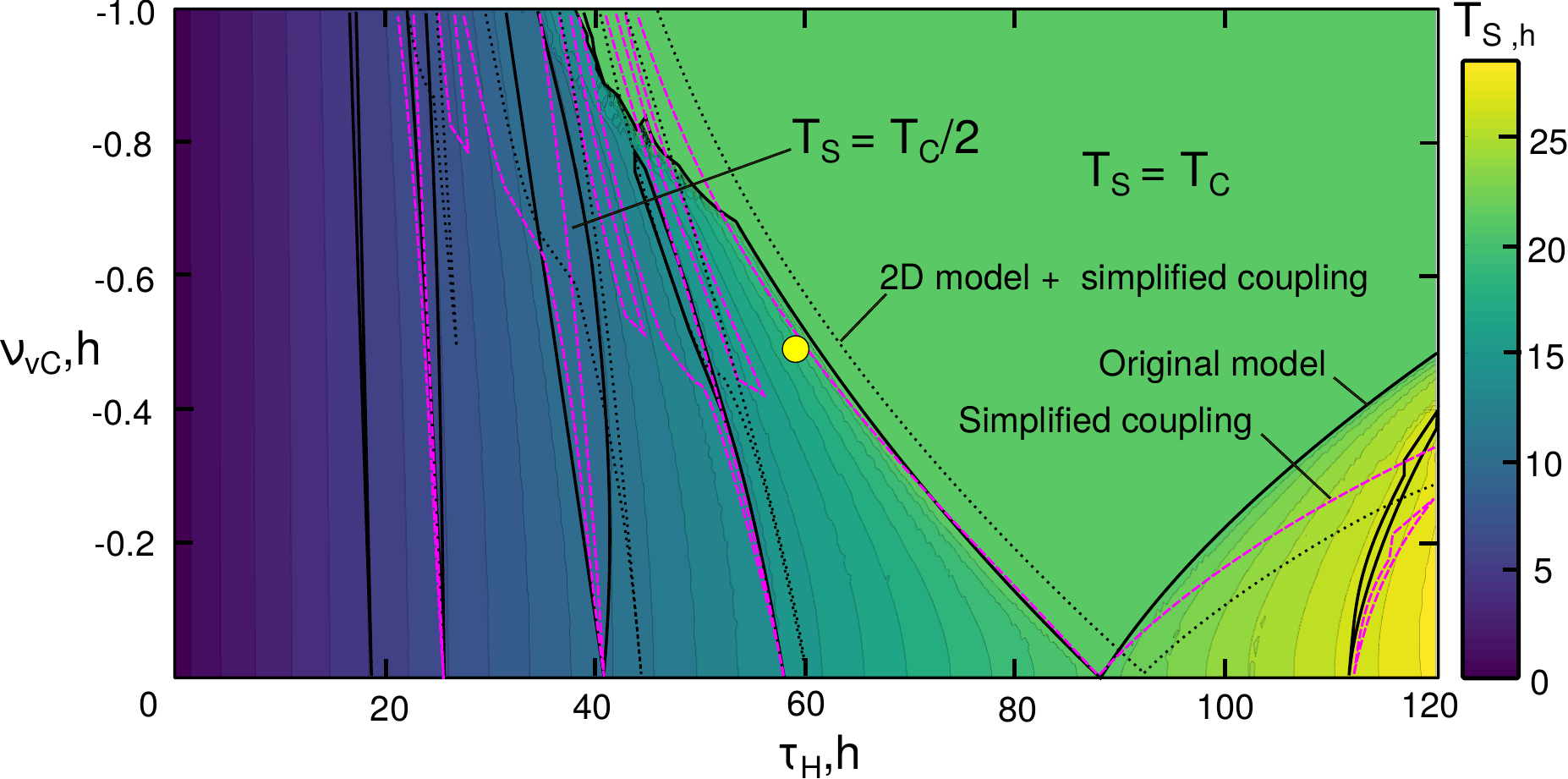}
\caption{Synchronisation map for the homeostatic oscillator in presence of one-directional coupling from the circadian oscillator. Both photic and non-photic inputs acting on the circadian oscillator are set to zero ($I_{ext}=0$ lx, $\nu_{Xn}=0$). All other parameters are at their default values. Colourbar corresponds to the period of the homeostatic oscillator (sleep period, $T_S$) and yellow filled circle shows the default parameter values, $\tau_H=59$ h and $\nu_{vC}=-0.5$ mV. The main resonances are shown with black and magenta lines for the three versions of the model: original model defined in Eqs (\ref{eqVv})-(\ref{eqI}), black lines; original model with simplified circadian coupling defined in Eq.~(\ref{approx}), magenta dashed lines; and reduced 2D model defined in Eqs(\ref{rAAm_1})-(\ref{rAAm_2}) with simplified circadian coupling  , black dotted lines.}
\label{arnold}
\end {center}
\end {figure} 

Figure \ref{arnold} shows the resulting synchronisation map for the homeostatic oscillator demonstrating $T_S$ at $\tau_h=0 \ldots 120$ h and $\nu_{vC}=0 \ldots -1$ mV. The period of the circadian oscillator $T_C$ is constant across the map and is equal to 24.13 h. This is because both photic and non-photic inputs are set to zero, so the circadian oscillator is not affected by the homeostatic oscillator or by light. Note that the period is different from the default value observed under constant darkness $T_C=24.2$ h set by $\tau_C$ because the circadian oscillator was calibrated in presence of the non-photic input, while here it is set to zero. 

Behaviour of the three model versions is compared: (i) original model of arousal dynamics, (ii) original model with simplified coupling, Eq.~(\ref{approx}), and (iii) simplified model with 2D homeostatic oscillator, Eqs~(\ref{rAAm_1}),(\ref{rAAm_2}) and simplified coupling. For all model versions, the main synchronisation region, where $T_S=T_C$, occupies the largest part of the diagram. For both the original model and the model with simplified coupling, the synchronisation tongue starts at $\tau_H=88.13$ h. For the model with 2D homeostatic oscillator, the tongue is shifted to $\tau_H=92$ h but its shape remains the same as seen by comparing the resonance lines (e.g., solid black and dashed magenta vs. dotted black). Interestingly, at $\nu_{vC}\in[-1, -0.5]$ mV, the model with simplified coupling produces higher number of synchronisation regions than the other two model versions as seen by the multiple tongues outlined by the dashed magenta lines in the top left region of the map. Importantly, however, the key synchronisation behaviour of the original model is conserved in simplified versions, especially for weak coupling, $\nu_{vC}\in[-0.5, 0]$ mV.  

Endogenous homeostatic period at the default value of $\tau_H=59$ h and $\nu_{vC}=0$ mV, is found to be $T_S=16.5$ h, which is $\approx 0.7T_C$. This is a significant difference in the oscillators periods, and, normally, synchronisation of two oscillators with such different periods requires either strong coupling strength or external driving force. However, amplitude of the coupling term $\nu_{vC}C(X,Y)$ is only 5\% of that for the variable $V_v$ on which it acts. It is, thus, not surprising that the default state of the model (yellow filled circle in Fig.~\ref{arnold}) is outside the main synchronisation range for all model versions in absence of the light-dark driving force and non-photic coupling. This means that the circadian and homeostatic oscillators in the original model at the default parameter values have tendency to be asynchronous in absence of external driving forces.
 
\subsection{Synchronisation of the three rhythms}
In this section we show that synchronisation of all three rhythms, the homeostatic, circadian and the light-dark cycle is only observed in a small range of $\tau_H$ values, while synchronisation of two of these three rhythms is more likely. 

\begin {figure} [htbp]
\begin {center}
\includegraphics [width=0.95\textwidth]{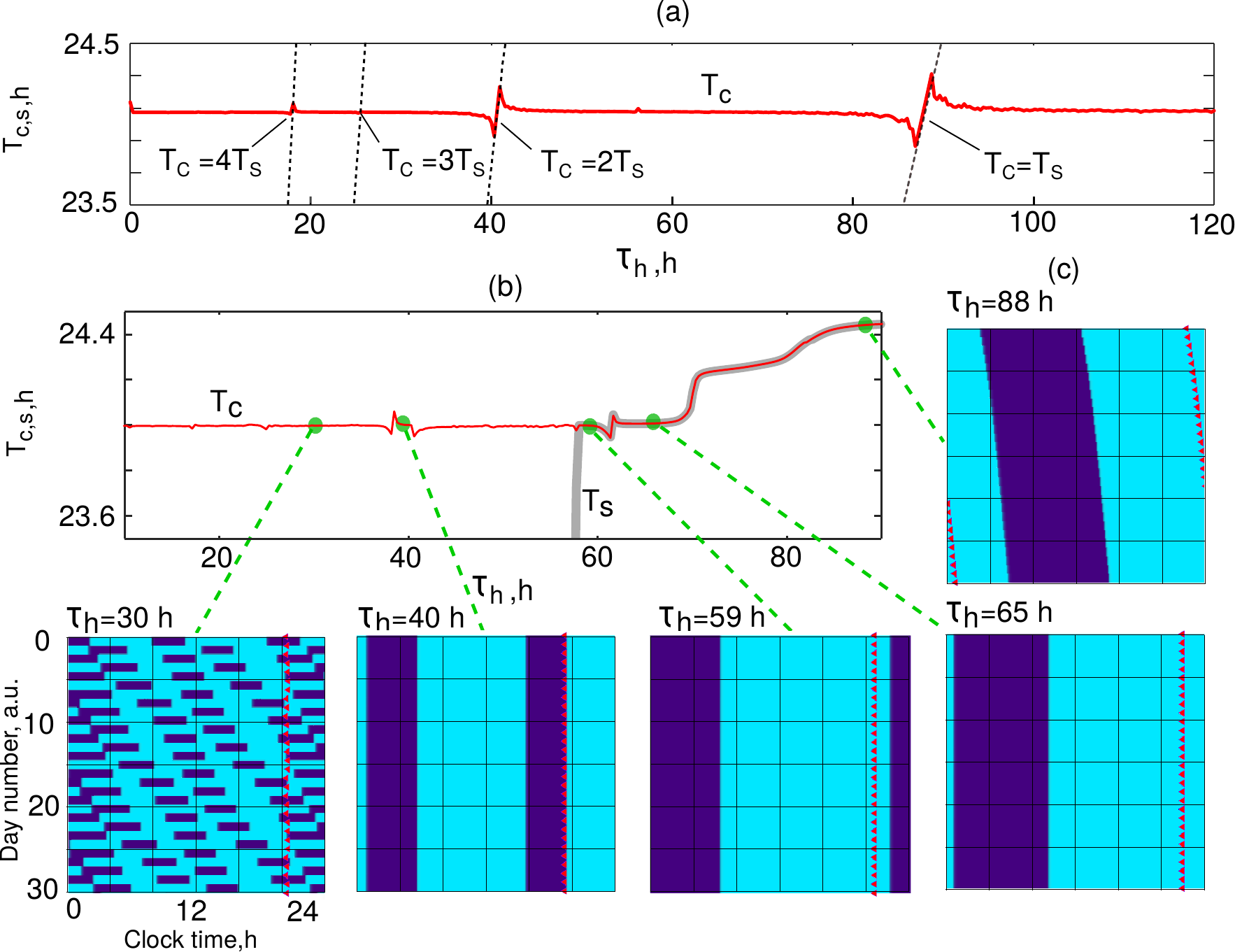}
\caption {Effects of non-photic coupling and light on synchronisation of the homeostatic and circadian oscillators. All results are for the original model. (a) Effect of $\tau_H$ on circadian period, $T_C$ (red line), and homeostatic period, $T_S$ (dashed black line), in presence on non-photic coupling. Circadian and photopic coupling are set to zero, $\nu_{vc}=0$ mV, $I_{ext}=0$ lx.(b) Dependence of $T_C$ and $T_S$ on $\tau_H$ in presence of all couplings and external light-dark cycle $I_{ext}=80$ lx. (d) Raster plots for selected examples of $\tau_H$ (values shown in panel titles) demonstrating 30-days dynamics of sleep times (blue lines) and circadian marker (onset of melatonin synthesis, red triangles) against clock time. Yellow indicates wakefulness. In all panels, simulations were run for 150 days. The periods presented in (a) and (b) are averaged over the last 100 days of the simulations, and the rasters in (c) are shown for the last 30 days. } 
\label{Ts_Tc}
\end {center}
\end {figure}

The non-photic coupling introduces disturbances in the periodicity of the circadian oscillator by dragging it away from the endogenous value of $\approx24.1$ h in small ranges of resonant values of $\tau_H$. This is seen in Fig.~\ref{Ts_Tc}(a) where $T_C$ and $T_S$ are shown for the case of default non-photic coupling while the circadian coupling and light input are set to zero ($\nu_{vC}=0$ mV, $I_{ext}=0$ lx). In the resonant areas where $T_C/T_S=1,2,4$ (but not where $T_C/T_S=3$), the circadian period follows the period of the homeostatic oscillator. This is characteristic behaviour for the phase/frequency locking mechanism of synchronisation \cite{balanov2008synchronization}. The non-photic coupling does not lead to large synchronisation areas as observed for the circadian coupling, but it is clear that it supports synchronisation of the homeostatic and circadian oscillators. 

Figure \ref{Ts_Tc}(b) shows similar calculations for $T_C$ and $T_S$ to those in Fig.~\ref{Ts_Tc}(a) but, this time, in the full model of arousal dynamics and in presence of all coupling terms and the external light-dark cycle, $I_{ext}=80$ lx, with a period of 24 h. In this case, the circadian oscillator is synchronised with the light dark cycle ($T_C=24$ h) but not the homeostatic oscillator ($T_S<<24$ h) at most values of $\tau_H<58.1$ h, except for small areas or resonance. The opposite is true for $\tau_H>69$ h, where the circadian and homeostatic oscillator are synchronised ($T_C=T_S$), but their periods are different from the 24-hour rhythm of the light-dark cycle and increase with the increase of $\tau_H$. Looking back at Fig.~\ref{arnold} where only circadian coupling is present, this range of $\tau_H$ at $\nu_{vC}=-0.5$ mV corresponds to large synchronisation zone of the homeostatic and circadian oscillators, but with fixed values of $T_S=T_C$ for $\tau_H\in[69,120]$. This difference in behaviour of the periods is due to the non-photic coupling slowing down the circadian period to follow the homeostatic one in Fig.~\ref{Ts_Tc}, while it was set to zero in Fig.~\ref{arnold}. 

Raster plots in Fig.~\ref{Ts_Tc}(c) show examples of sleep-wake cycles and timing of melatonin synthesis onset (experimentally used marker of the circadian phase) observed over 30 days at selected values of $\tau_H$. At $\tau_H=30$ h, $T_C/T_S\approx 3$ and there are 3-4 short sleep episodes every day with their position changing daily, while melatonin markers remain fixed at period of 24 h. At $\tau_H=40$ h there is a resonance with $T_C/T_S\approx2$ and the sleep-wake cycles are synchronised with the 24 h rhythm of the circadian oscillator and the light dark cycle but there are two sleep episodes per day and melatonin synthesis onset appears at the end of the second sleep episode. At $\tau_H=59$ h, regular sleep-wake cycles are observed with one 8-hour sleep episode per day starting at $\approx$ 22:00. These are accompanied by the 24 h rhythm of melatonin markers which appear about one hour before the sleep onset. Alltogether, this corresponds to a typical sleep and circadian pattern for healthy people \cite{Abeysuriya18}. Further increase of $\tau_H$ to 65 h results in the shift of the sleep episodes to later time while the timing of melatonin onset shifts only slightly resulting in a larger time gap between melatonin onset and sleep start. However, all the rhythms remain synchronised. At $\tau_H=88$ h, the sleep wake cycles are synchronised with the circadian oscillator (phase difference between sleep and melatonin onset is constant) but both are different from the 24 h period of the light-dark cycle.   
 
\subsection{Effects of external neuronal inputs to the MA and VLPO on synchronisation}  
Previous sections focused on the role of $\tau_H$ and coupling terms in synchronisation. However, position of nullclines in Fig.~\ref{nulls} and oscillators properties also depend on other parameters. In particular, parameters $A_m$ and $A_v$ represent external neuronal inputs to the homeostatic oscillator from other brain nuclei and are likely to be varied depending on an individual and their physical or mental state. In this section we study how changes in $A_m$ and $A_v$ affect the dynamics.
First, we do it for the blocked coupling from the circadian oscillator, at $\nu_{vC}=0$, in order to see reveal the intrinsic features of the homeostatic oscillator.
Second, we do it in the full model with circadian and non-photic couplings but in absence of driving force ($I_{ext}=0$). It allows one to assess the contribution from reciprocal coupling between two oscillators.  

\begin {figure} [htbp]
\begin {center}
 \includegraphics[width=0.9\textwidth]{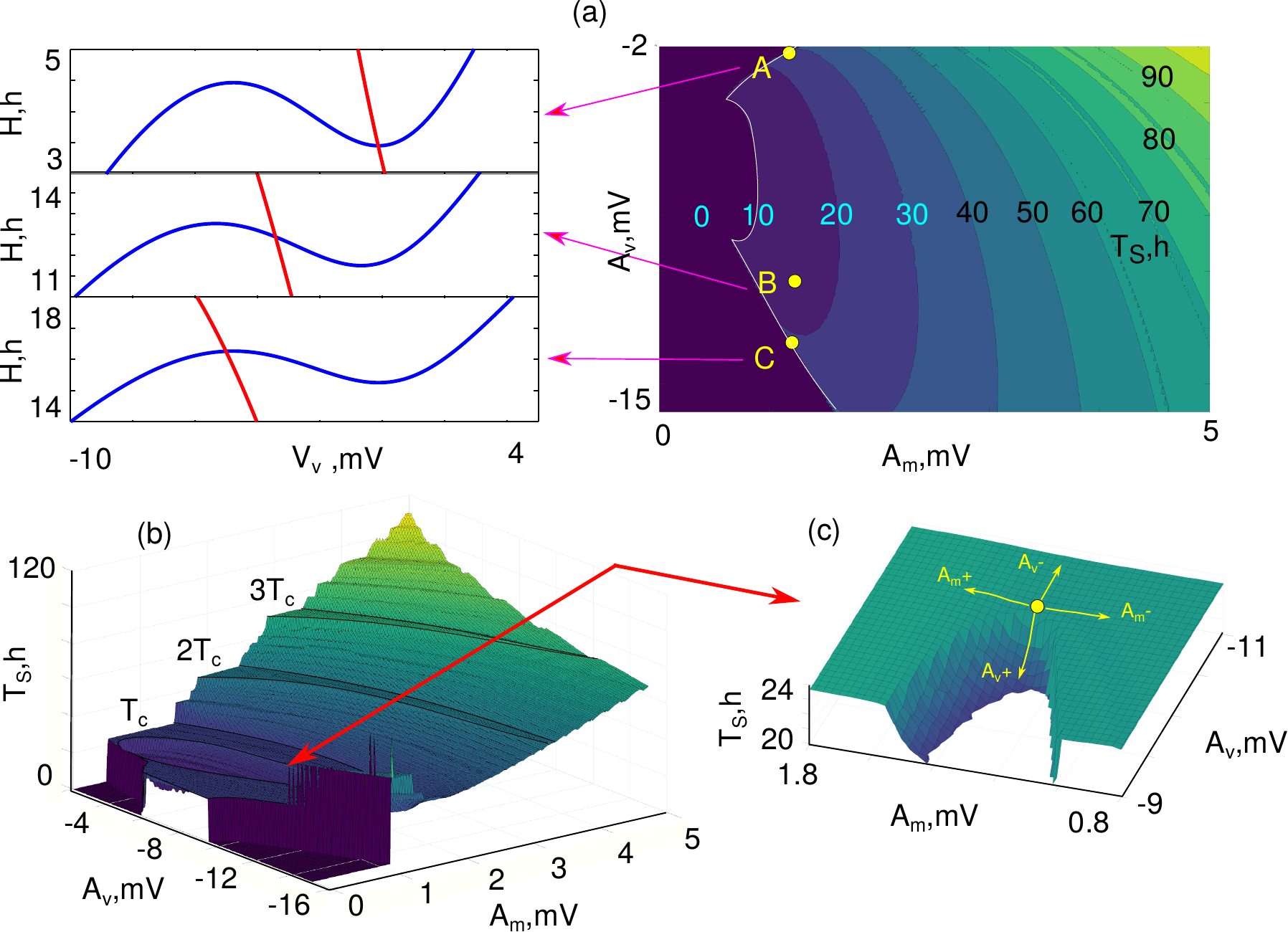}
\caption {Effects of $A_m$ and $A_v$ on the homeostatic period, $T_S$. (a) Map of periods for the homeostatic oscillator, $T_S$ (right) and nullclines (left)  for representative points A, B, and C in the map in absence of circadian coupling, $\nu_{vC}=0$. Colourcoding in the map indicates different values of $T_S$ with the contours indicating lines of equal $T_S$. Numbers indicate relevant $T_S$ values in hours. $H$-nullcline is shown in red and $V_v$-nullcline in blue. (b) Map of $T_S$ in presence of default circadian and non-photic coupling, but zero light, $I_{ext}=0$ lx.(c) Zoomed-in area of (b) showing location of the default values of $A_m=1.3$ mV and $A_v=-10.3$ mV. Arrows show direction of increase/decrease of the parameters.}  
\label{Am_Av}
\end {center}
\end {figure}

Figure \ref{Am_Av}(a)-right shows response of $T_S$ to changes in $A_m$ and $A_v$ in absence of circadian input to the homeostatic oscillator. As seen from the direction of the contour lines in the map, the change of $A_m$ has stronger effect on $T_S$ than the change of $A_v$. However, at the default parameter values (point B) the homeostatic oscillator is sensitive to both parameters. Decrease of $A_v$ starting at point B leads to slowing of the oscillations until they disappear completely at point C. Increase of $A_v$ leads to similar behaviour and disappearance of oscillations at point A. These dynamics are explained by the nullclines in Fig.~\ref{Am_Av}(a)-left. For both points A and C the $H-$ and $V_v-$nullclines cross at one of the extrema of the $V_v-$nullcline. This situation is well-studied in FHN model and corresponds to a transition from self-sustained oscillations to excitable dynamics  \cite{fitzhugh1961impulses,izhikevich2007dynamical}. Mathematically, this corresponds to presence of the supercritical Andronov-Hopf bifurcation with the so-called Canard explosion \cite{krupa2001relaxation}. Thus oscillations disappear at $A_m$ values to the left of points A and C, and more generally to the left of the yellow line in the map. 

The $T_S$ map changes significantly when circadian coupling is set to its default value, $\nu_{vC}=-0.5$ mV ($I_{ext}$ is still zero). In this case, the map shows a step-like structure where $T_S$ lingers at fixed resonant values of $T_S=nT_C$, where $n=1,2,3...$, while $A_m$ and $A_v$ are changed until it transitions to the next resonance, Fig.~\ref{Am_Av}(b).
This set of frequency/phase-locked states makes the effect of $A_v$, and especially, $A_m$ changes highly dependent on its specific choice: from negligibly weak within the resonant areas to abrupt changes at their borders. 

Similar to Fig.~\ref{Am_Av}(a), oscillations disappear at low $A_m$ and increase of $A_m$ leads to higher $T_S$, while $A_v$ has minor effect on $T_S$. The location of the default values of $A_m$ and $A_v$ is shown in Fig.~\ref{Am_Av}(c),  which makes it clear that in the default state the model is more sensitive to small decrease of $A_v$ than changes of $A_m$ or increase of $A_v$. With the default parameter values sitting on the border of the $T_S=T_C$ resonance, decrease of $A_v$ would further desynchronise the system.

\section{Discussion}
We have applied methods of nonlinear dynamics to study synchronisation in the model of arousal dynamics and showed that the key model element, the 3D homeostatic oscillator can be simplified to 2D form without significant change in its synchronisation properties. By approximating one of the nullclines of the 2D oscillator with a line, we showed that the 2D homeostatic oscillator is equivalent to the well-known Fitzhugh-Nagumo model, which is a widely used model in nonlinear physics \cite{fitzhugh1961impulses,izhikevich2007dynamical}. The circadian oscillator, on the other hand, is modelled with the Van der Pol oscillator \cite{StHilaire07}, the only difference from the classical Van der Pol oscillator \cite{van1926lxxxviii,kanamaru2007van} is the higher degree of polynomial used to describe variable $X$ - seven instead of three. Taking into consideration the coupling between the two oscillator, the full model can thus be described as a periodically forced and reciprocally coupled two 2D self-sustained oscillators, one showing  smooth oscillations, and another -  fast-slow dynamics. Mathematically, such system is represented by trajectories moving on a 3D torus
\cite{ruelle1971nature,yoshimoto1993coupling}, and the resonances between the homeostatic, circadian and light-dark cycles fit well with this paradigm \cite{anishchenko2008bifurcational}.  

\subsection{Implications for physiology and modelling of sleep-wake cycles}
The default state of the model of arousal dynamics represents sleep-wake and circadian dynamics for a typical healthy young individual (or a group average). The model has been validated against $>50$ experimental datasets and successfully reproduces variety of sleep phenomena \cite{postnova2016sleep,Postnova18,Abeysuriya18}. From synchronisation point of view, the default state of the model corresponds to the global resonance 1:1:1. Interestingly, however, this state is far from parameter values where the homeostatic oscillator has a period close to the circadian oscillator and the light-dark cycle. Instead the default state of the model is located at the border of synchronisation range, where, in absence of the circadian input, the endogenous homeostatic period is $T_S\approx16$ h. In this uncoupled state multiple short ($<<8$ h) sleep episodes occur per day with a mean total daily sleep being longer than under normal synchronised conditions (11.25 h with no circadian input vs. 8.5 h with intact circadian coupling). Importantly, these findings are in line with experimental data showing that in SCN-lesioned squirrel monkeys, sleep bouts are shorter but the total daily sleep duration is longer than in intact animals \cite{Edgar93}.

The fact that in absence of light but with intact circadian coupling, the model functions at the border of the synchronous regime means that it is very close to the so-called spontaneous internal desynchrony (SID) - a phenomenon that was experimentally observed in people living for extended periods of time in constant darkness. During SID the circadian oscillator usually has period close to 24 h, whereas sleep appears with much shorter (12-20 h) or longer (28-68 h) period \cite{Aschoff67,Aschoff71,Wever75,Wever79,Phillips11,Gleit13} - similar to the dynamics seen in our study. This ease of transition from normal sleep-wake cycles under the 24 h light-dark cycles to SID under darkness in humans indicates that the brain operates close to a bifurcation point and only presence of environmental driving force allows it to have stable and synchronised periodic activity. The model of arousal dynamics was not originally designed to reproduce SID, but its ability to do so with only a small nudge towards lower $\tau_H$ and/or lower $\nu_{vC}$ supports the current set of default parameters. If the model default state was deep in the synchronous regime, it would have been difficult to achieve SID with physiologically justified parameter changes. 

Our results predict two main types of desynchrony in the model. First is at $\tau_H\in[0,58]$ h, where the circadian oscillator with a typical, experimentally confirmed \cite{Czeisler99}, endogenous period of 24.1-24.2 h is entrained to the 24 h light dark cycle, but the homeostatic oscillator is asynchronous. In this case, melatonin synthesis onset appears at a fixed time every day and at the correct time of day (physiological range between 19:00 to 01:00 \cite{Sletten15}) but sleep times are not phase locked with it. From the appearance of sleep-wake cycles and melatonin rhythms, especially at non-resonant values of $\tau_H$, this regime is easily confused with the one where circadian coupling is zero. The mechanism, however, is different because  the circadian coupling is intact but it is the homeostatic time constant that causes the desynchrony. Second, is at $\tau_H>69$ h where the homeostatic and circadian oscillators are synchronised $T_S=T_C$ but are different from the 24 h of the light-dark cycle. In this case the onsets of melatonin synthesis and of sleep are phase locked in correct relationship (melatonin preceding sleep) but the phase angle increases with increase of $\tau_H$ and both melatonin and sleep shift to later and later time every day. 

\subsection{Insights into potential mechanisms of sleep and circadian disorders}
The desynchronisation patterns discussed above may be linked to those observed in circadian rhythm sleep disorders \cite{Sack07}. These include diseases where (i) sleep appears several hours earlier than conventional or desired sleep time - this is known as advanced phase sleep disorder, ASPD; (ii) sleep appears substantially later than the conventional sleep time - delayed sleep phase disorder, DSPD; and (iii) sleep does not follow circadian rhythms and instead of being consolidated into a single episode per day, there are several shorter sleep episodes that appear at random times throughout day and night - this is knowm as irregular sleep-wake rhythm, ISWR \cite{Sack07}. In all these diseases, it is challenging (sometimes impossible) to maintain socially-conventional schedules of work and commitments and results in further disturbances of sleep and health \cite{Rajaratnam15}. All these diseases are generally ascribed to disturbances in the circadian system but their exact mechanisms are unknown \cite{Sack07}. Our study shows that all these circadian sleep patterns can also be obtained by changing $\tau_H$ while the circadian system remains unchanged. We predict that ISWR dynamics would be observed at very short, non-resonant $\tau_H$, e.g., $\tau_H<40$ h, ASPD would be seen at moderatly short values of $\tau_H$ just below the default state, and DSPD dynamics would be seen at $\tau_H>69$ h. Similarly, it is commonly assumed that timing of melatonin marker is caused by changes in the circadian system. Our study shows, that change in $\tau_H$ can cause advance and delay of melatonin onset relative to clock time and to sleep onset, without any changes in the cicradian oscillator itself. These are critical insights which may help us better understand mechanisms of these diseases and develop better diagnosis and treatment procedures. Future research should thus focus on both the circadian- and homeostatic-driven pathways to these diseases. 

Finally, we showed that parameters $A_m$ and $A_v$ responsible for inputs from other neuronal populations to the MA and the VLPO also affect the period of the homeostatic oscillator and synchronisation. This means that these other populations, e.g., orexin neurons, may likewise be involved in either support of synchronisation or desynchronisation of the homeostatic and circadian oscillators and the light-dark cycle. In practical sense it means that information on sleep-wake patterns and melatonin timing during circadian rhythms sleep disorders is insufficient for understanding which system components led to desynchronisation. Interestingly, low values of $A_m$ lead to complete loss of self-sustained oscillations - a situation that is not observed at variation of the other parameters studied here. In real life such complete loss of sleep is observed in fatal familial insomnia - a rare genetic neurodegenerative disorder characterised by complete inability to sleep and loss of some autonomic functions (e.g., temperature control), which has no cure and ultimately leads to death \cite{Goldfarb92}. Our study predict that it may be caused by degeneration of neuronal populations acting on the MA, which should be studied further in the future.


\subsection{Study limitations and future work}
In this study we have considered only two light profiles: one is a constant darkness with $I_{ext}=0$ lx and the other is the 12/12 light-dark cycle with $I_{ext}=80$ lx resulting in darkness at night between 20:00 and 08:00 and sinusoidally modulated light during daytime (08:00 to 20:00) peaking at 80 lx. Intensity of light and shape of light profile (timing) strongly affect the dynamics of the circadian oscillator \cite{Golombek10}, which has been extensively studied in circadian models, e.g. \cite{Tekieh2020}. However, systematic studies of how light intensity and timing affect synchronisation of all three rhythms are lacking and need to be performed in the future. This will aid in better understanding of dynamics and design of optimal light schedules for such common examples of circadian misalignment like shiftwork and jetlag. 

Majority of sleep models, including the model of arousal dynamics, are deterministic. To ensure a stable state of the model at different parameter values we had to perform simulations of hundreds of days of sleep-wake cycles. However, in real life our sleep-wake cycles are affected by numerous random processes and neither sleep times nor melatonin markers are identical from day to day. Mathematically, this means that stochastic model of sleep-wake cycles needs to be implemented accounting for different sources of randomness, both endogenous (e.g., fluctuations in dynamics of neuronal populations \cite{takahashi2010locus}) and external (e.g., changes of light, stress, meals, exercise \cite{Logan19}). Addition of these endogenous dynamics, in particular those of neuronal populations will allow to bridge the gap between the current models of sleep-wake cycles and brain clearance mechanisms during sleep, which has recently been shown to be under control of some of the same populations as implemented in the model of arousal dynamics \cite{o2015distinct}.

\section*{Acknowledgments}
This research was supported by   the Russian Ministry of Education and Science, project \#075-15-2019-1885

\section*{Conflicts of Interest}
DEP, KOM, and SP have no conflicting interests to declare. In interest of full disclosure: SP served as a Theme Leader and previously as a Project Leader in the CRC for Alertness, Safety and Productivity which funded development of the model of arousal dynamics. She reports research grants from Qantas Airways Ltd and Alertness CRC, which are not related to this paper.

\bibliographystyle{unsrt}

\end{document}